\documentclass[11pt]{article}
\usepackage{t1enc}
\usepackage[utf8]{inputenc}
\usepackage[english]{babel}
\usepackage{graphicx}
\usepackage{amssymb}
\usepackage{amsmath}
\usepackage{amsthm}

\def\OA{{\cal A}}

\def\SDK{{\cal K}}

\def\ST{{\cal M}}

\def\N{{\mathcal N}}

\def\RE{{\bf R}}

\def\A{{\cal A}}
\def\B{{\cal B}}

\def\RE{{\mathbb R}}

\def\qed{\ \vrule height 5pt width 5pt depth 0pt}
\def\cros{\raise1.9pt\hbox{$\scriptscriptstyle
          >$}\!\raise1.5pt\hbox{$\scriptstyle\triangleleft\,$}}

\def\C{{\mathcal C}}

\def\T{{\mathcal T}}
\def\V{{\mathcal V}}
\def\G{{\mathcal G}}

\def\cL{{\cal L}}
\def\S{\Sigma}
\def\s{\sigma}

\def\N{{\cal N}}
\def\B{{\cal B}}

\theoremstyle{definition}\newtheorem{D}{Definition}
\theoremstyle{definition}\newtheorem{Prop}{Proposition}
\theoremstyle{definition}
\newcommand{\noi}{\vspace{0.1in} \noindent}

\title{\bf Bell's local causality is a d-separation criterion}
\author{\textit{G\'abor Hofer-Szabó}\thanks{Research Center for the Humanities, Budapest; email: szabo.gabor@btk.mta.hu}}
\date{ }
\begin{document}
\maketitle

\begin{abstract}
This paper aims to motivate Bell's notion of local causality by means of Bayesian networks. In a locally causal theory any superluminal correlation should be screened off by atomic events localized in any so-called \textit{shielder-off region} in the past of one of the correlating events. In a Bayesian network any correlation between non-descendant random variables are screened off by any so-called \textit{d-separating set} of variables. We will argue that the shielder-off regions in the definition of local causality conform in a well defined sense to the d-separating sets in Bayesian networks.
\vspace{0.1in}

\end{abstract}

\section{Introduction}

\begin{figure}[ht]
\centerline{\resizebox{14cm}{!}{\includegraphics{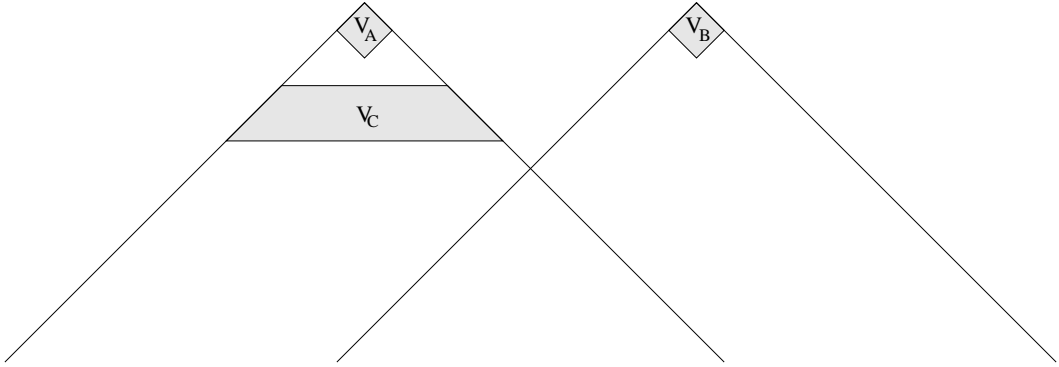}}}
\caption{Full specification of what happens in $V_C$ makes events in $V_B$ irrelevant for predictions about $V_A$ in a locally causal theory.}
\label{dsep1}
\end{figure}

John Bell's notion of local causality is one of the central notions in the foundations of relativistic quantum physics. Bell himself has returned to the notion of local causality from time to time providing a more and more refined formulation for it. The final formulation stems from Bell's posthumously published paper ``La nouvelle cuisine.'' It reads as follows:\footnote{For the sake of uniformity we slightly changed Bell's notation and figure.}
\begin{quote}
``A theory will be said to be locally causal if the probabilities attached to values of local beables in a space-time region $V_A$ are unaltered by specification of values of local beables in a space-like separated region $V_B$, when what happens in the backward light cone of $V_A$ is already sufficiently specified, for example by a full specification of local beables in a space-time region $V_C$.''  (Bell, 1990/2004, p. 239-240)
\end{quote}
The figure Bell is attaching to his formulation of local causality is reproduced in Fig.\ \ref{dsep1} with Bell's original caption. 
In a rough translation, a theory is locally causal if any superluminal correlation can be screened-off by a ``full specification of local beables in a space-time region'' in the past of one of the correlating events.

The terms in quotation marks, however, need clarification. What are ``local beables''? What is ``full specification'' and why is it important? Which are those regions in spacetime which, if fully specified, render superluminally correlating events probabilistically independent? The first two questions have attracted much interest among philosophers of science. As Bell puts it, ``\textit{be}ables of the theory are those entities in it which are, at least tentatively, to be taken seriously, as corresponding to something real'' (Bell, 1990/2004, p. 234). Furthermore, ``it is important that events in $V_C$ be specified completely. Otherwise the traces in region $V_B$ of causes of events in $V_A$ could well supplement whatever else was being used for calculating probabilities about $V_A$''  (Bell, 1990/2004, p. 240).

\begin{figure}[ht]
\centerline{\resizebox{14cm}{!}{\includegraphics{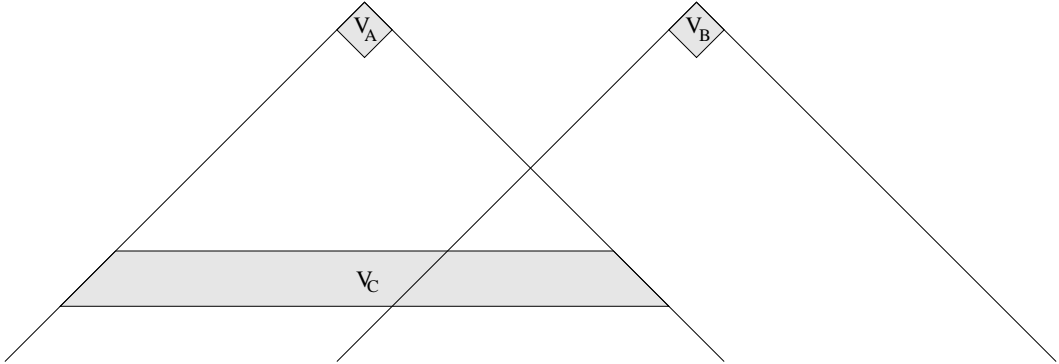}}}
\caption{A \textit{not} completely shielding-off region $V_C$.}
\label{dsep2}
\end{figure}

The third question, however, concerning the localization of the screener-off regions has gained much less attention in the literature. How to characterize the regions which region $V_C$ in Fig.\ \ref{dsep1} is an example of? Bell's answer is instructive but brief: ``It is important that region $V_C$ completely shields off from $V_A$ the overlap of the backward light cones of $V_A$ and $V_B$.''  (Bell, 1990/2004, p. 240) But why to shield off the common past of the correlating events? Why the region $V_C$ cannot be in the remote past of $V_A$ as for example in Figure \ref{dsep2}? 
Well, intuition dictates that in this latter case some event might occur \textit{above} the shielder-off region but still \textit{within} the common past establishing a correlation between events in $V_A$ and $V_B$. This intuition is correct. The aim of this paper, however, is to provide a more precise explanation for the localization of the shielder-off regions in spacetime. This explanation will consists in drawing a parallel between local physical theories and Bayesian networks. It will turn out that \textit{the shielder-off regions in the definition of local causality play an analogous role to the so-called d-separating sets of random variables in Bayesian networks}. 

\noi
There is a renewed interest in Bell's notion of local causality (Norsen, 2009, 2011; Maudlin 2014), its relation to separability (Henson, 2013b); the role of full specification in local causality (Seevinck and Uffink, 2011; Hofer-Szabó 2015a); its role in relativistic causality (Butterfield 2007; Earman and Valente, 2014; Rédei 2014); its status as a local causality principle (Henson, 2005; Rédei and San Pedro, 2012; Henson 2013a). A similar closely related topic, the Common Cause Principle is also given much attention (Rédei 1997; Rédei and Summers 2002; Hofer-Szabó and Vecsernyés 2012a, 2013a). On the other hand, there is also an intensive discussion on the applicability of the Causal Markov Condition in the EPR scenario (Glymour, 2006; Suárez and Iniaki, 2011; Hausman and Woodward, 1999; Suárez, 2013; Hofer-Szabó, Rédei and Szabó, 2013). Despite the rich and growing literature on the topic I am unaware of any work relating Bayesian networks and especially d-separation directly to local causality. This paper intends to fill this gap. For a precursor of this paper investigating Causal Markov Condition in a specific local physical theory see (Hofer-Szabó, 2015b).

In the paper we will proceed as follows. In Section 2 we introduce the basics of the theory of Bayesian networks and the notion of d-separation and m-separation. In Section 3 we define the notion of a local physical theory and formulate Bell's notion of local causality within this framework. We prove our main claim in Section 4 and conclude in Section 5.

\section{Bayesian networks and d-separation}

A \textit{Bayesian network} (Pearl, 2000; Glymour, Scheines and Spirtes, 2000) is a pair $(\G, \V)$ where $\G$ is a directed acyclic graph and $\V$ is a set of random variables on a classical probability space $(X, \S,p)$ such that the elements $A, B \dots$ of $\V$ are represented by the vertices of $\G$ and the arrows (directed edges) $A\rightarrow B$ on the graph represent that $A$ is \textit{causally relevant} for $B$. Two vertices are called \textit{adjacent} if they are connected by an arrow. For a given $A\in\V$, the set of vertices that have directed edges in $A$ is called the \textit{parents} of $A$, denoted by $Par(A)$; the set of vertices from which a directed paths is leading to $A$ is called the \textit{ancestors} of $A$, denoted by $Anc(A)$; and finally the set of vertices that are endpoints of a directed paths from $A$ is called the \textit{descendants} of $A$, denoted by $Des(A)$. For a set $\C$ of vertices $Par(\C)$, $Anc(\C)$ and $Des(\C)$ are defined similarly. 

The set $\V$ is said to satisfy the \textit{Causal Markov Condition} \index{Causal Markov Condition} relative to the graph $\G$ if for any $A\in\V$ and any $B\notin Des(A)$ the following is true:
\begin{eqnarray}\label{CMC}
p(A \, | \, Par(A) \wedge B) & = & p(A \, | \, Par(A))
\end{eqnarray}
or equivalently
\begin{eqnarray}\label{CMC2}
p(A \wedge B \, | \, Par(A)) & = & p(A \, | \, Par(A)) \, p(B \, | \, Par(A))
\end{eqnarray}
That is conditioning on its parents any random variable will be probabilistically independent from any of its non-descendant. Non-descendants can be of two types: either ancestors or \textit{collaterals} (non-descendants and non-ancestors). As we will see, being independent of collaterals is what relates the Causal Markov Condition to Bell's local causality. 

Causal Markov Condition establishes a special conditional independence relation between some random variables of $\V$. But there are many other conditional independences. In a faithful Bayesian network these other conditional independences are all implied by the Causal Markov Condition by means of the so-called \textit{d-separation} criterion. Let $\mathcal{P}$ be a \textit{path} in $\G$, that is a sequence of adjacent vertices. A variable $E$ on $\mathcal{P}$ is a \textit{collider} if there are arrows to $E$ from both its neighbors on $\mathcal{P}$ ($D \rightarrow E \leftarrow F$). Now, let $\C$ be a set of vertices and let $A$ and $B$ two different vertices not in $\C$. The vertices $A$ and $B$ are said to be \textit{d-connected} by $\C$ in $\G$ iff there exists a path $\mathcal{P}$ between $A$ and $B$ such that every non-collider on $\mathcal{P}$ is not in $\C$ and every collider is in $Anc(\C)$. $A$ and $B$ are said to be \textit{d-separated} by $\C$ in $\G$, iff they are not d-connected by $\C$ in $\G$.

The intuition behind d-separation is the following. A vertex $E$ on a path (not at the endpoints) can be either a \textit{collider} ($D \rightarrow E \leftarrow F$), an \textit{intermediary cause} ($D \rightarrow E \rightarrow F$) or a \textit{common cause} ($D \leftarrow E \rightarrow F$). The idea here is that only intermediary and common causes (together called \textit{non-colliders}) can transmit causal dependence and hence establish probabilistic dependence. This dependence can be \textit{blocked} by conditioning on the non-collider. Colliders behave just the opposite way. They represent two events causing a common effect. These two causes are causally and probabilistically independent, but become dependent upon conditioning on their common effect. Moreover, they also become dependent upon conditioning on any of the descendants of the effect. Putting these together, the causal dependence on a path $\mathcal{P}$ connecting two vertices is blocked by a set $\C$ if either there is at least one non-collider on $\mathcal{P}$ which is in $\C$ or there is at least one collider $E$ on $\mathcal{P}$ such that either $E$ or a descendant of $E$ is not in $\C$. The two vertices are d-separated by $\C$ if causal dependence is blocked on every path connecting them. 

\begin{figure}[ht]
\centerline{\resizebox{14cm}{!}{\includegraphics{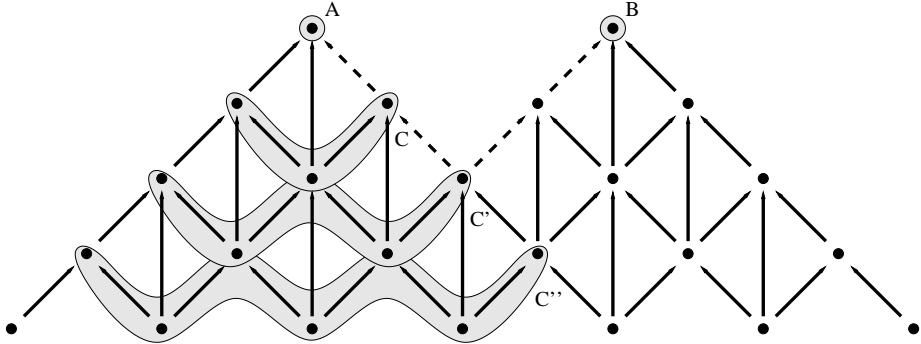}}}
\caption{$A$ and $B$ are d-separated by $\C$ and $\mathcal{C'}$ but d-connected by $\mathcal{C''}$.}
\label{dsep4}
\end{figure}

As an example for d-connection and d-separation consider the causal graph in Fig.\ \ref{dsep4}.
(The arrows are directed to up, left up and right up.) Let $A$ be the left ``peak'' and $B$ the right ``peak'' in the graph and let $\C$, $\mathcal{C'}$ and $\mathcal{C''}$ be the sets shown in the figure containing 3, 5 and 7 vertices, respectively. Then $A$ and $B$ are d-separated by $\C$ since the parents are always d-separating due to the Causal Markov Condition. $A$ and $B$ are d-separated also by $\mathcal{C'}$ since for every path connecting the peaks there is a non-collider in $\mathcal{C'}$. However, $A$ and $B$ are d-connected by $\mathcal{C''}$ since there is a path (denoted by a broken line in Fig.\ \ref{dsep4}) connecting the peaks which contains only non-colliders outside $\C''$. Consequently, the following probabilistic relations hold:
\begin{eqnarray}
p(A \wedge B \, | \, \C) & = & p(A \, | \, \C) \, p(B \, | \, \C) \label{dsepC} \\
p(A \wedge B \, | \, \mathcal{C'}) & = & p(A \, | \, \mathcal{C'}) \, p(B \, | \, \mathcal{C'}) \label{dsepC'} \\
p(A \wedge B \, | \, \mathcal{C''}) & \neq & p(A \, | \, \mathcal{C''}) \, p(B \, | \, \mathcal{C''}) \label{dsepC''}
\end{eqnarray}

Looking at in Fig.\ \ref{dsep4}, what stands out immediately is that a set which is too far in the causal past of $A$ cannot d-separate $A$ from a collateral event since there might be paths connecting them ``above''  the set. As we will see, a similar moral will be valid in case of local causality: regions with are too far in the causal past of an event cannot screen it off from a spacelike separated event since there might be events ``above'' the region which can establish correlation between them.

\noi
In analyzing local causality sometimes we need to go beyond directed acyclic graphs. A graph which may contain both directed ($A\rightarrow B$) and bi-directed ($A\leftrightarrow B$) edges is called \textit{mixed}. The d-separation criterion extended to mixed acyclic graphs is called \textit{m-separation}. (Richardson and Spirtes, 2002; Sadeghi and Lauritzen, 2014) Two vertices $A$ and $B$ are said to be \textit{m-connected} by $\C$ in a mixed acyclic graph $\G$ iff there exists a path $\mathcal{P}$ between $A$ and $B$ such that every non-collider on $\mathcal{P}$ is not in $\C$ and every collider is in $Anc(\C)$. $A$ and $B$ are said to be \textit{m-separated} by $\C$ in $\G$, iff they are not m-connected by $\C$ in $\G$. In a directed acyclic graph m-separation reduces to d-separation.

\begin{figure}[ht]
\centerline{\resizebox{14cm}{!}{\includegraphics{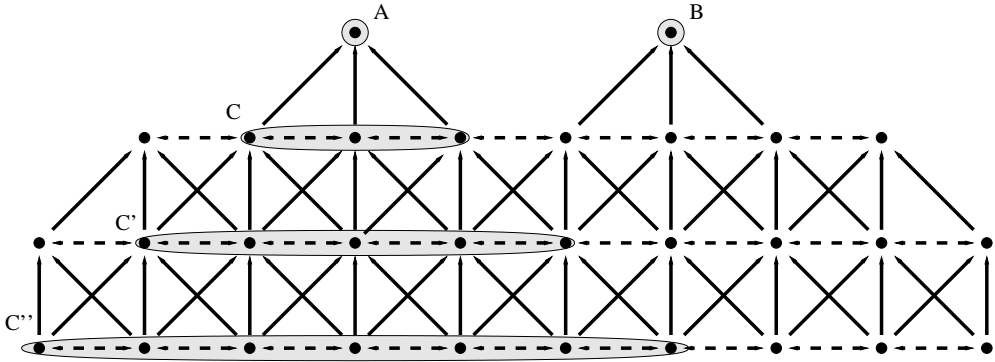}}}
\caption{$A$ and $B$ are m-separated by $\C$ but m-connected by both $\mathcal{C'}$ and $\mathcal{C''}$.}
\label{dsep5}
\end{figure}

An example for a mixed acyclic graph is depicted in Fig.\ \ref{dsep5}.
Here the bi-directed edges are represented by dotted lines. Again, let $A$ be the left ``peak'' and $B$ the right ``peak'' in the graph and let $\C$, $\mathcal{C'}$ and $\mathcal{C''}$ be the sets shown in the figure containing 3, 5 and 7 vertices, respectively. Then $A$ and $B$ are m-separated by $\C$ but m-connected by both $\mathcal{C'}$ and $\mathcal{C''}$. The connecting path is the shortest path connecting $A$ and $B$.

\noi
Now, let us connect the terminology of Bayesian networks to that of standard physics. A random variable is a real-valued  Borel-measurable function on $X$. Each random variable $A \in \V$ generates a sub-$\sigma$-algebra of $\S$ by the inverse image of the Borel sets:
\begin{eqnarray}
\s(A) := \left\{ A^{-1} (b) \, | \, b \in \B(\RE) \right\}
\end{eqnarray}
Similarly, each set $\C$ of $n$ random variables generates a sub-$\sigma$-algebra of $\S$ by the inverse image of the $n$-dimensional Borel sets:
\begin{eqnarray}
\s(\C) := \left\{ (C_1, C_2 \dots C_n)^{-1} (b) \, | \, C_i \in \C, \, \, b \in \B(\RE^n) \right\}
\end{eqnarray}
From this perspective d-separation tells us which sub-$\s$-algebras are probabilistically independent conditioned on which other sub-$\s$-algebras of $\S$.

Now, instead of using $\s$-algebras it is more instructive to use a richer structure in physics, namely \textit{von Neumann algebras}. Consider the characteristic functions on $X$ projecting on the elements of $\S$, called \textit{events}. The set $\{\chi_S\, | \, S \in \S\}$ of characteristic functions generates an abelian von Neumann algebra, namely $\cL^\infty(X,\S,p)$, the space of essentially bounded complex-valued functions on $X$. Starting from the characteristic functions of the sub-$\s$-algebra $\s(A)$, one arrives at a subalgebra of $\cL^\infty(X,\S,p)$. Denote this abelian von Neumann algebra determined by the random variable $A$ by $\N_A$. Similarly, denote by $\N_\C$ the von Neumann algebra determined by a set $\C$ of random variables. 

Instead of using a probability measure on $\S$ or on a sub-$\sigma$-algebra $\s(A)$, one can also use a state on the corresponding von Neumann algebra $\N_A$. A \textit{state} $\phi$ is a positive linear functional of norm 1 on a von Neumann algebra. States on $\N_A$ and probability measures on $\s(A)$ mutually determine one another: a state restricted to the characteristic functions in $\N_A$ is a probability measure on $\s(A)$; and vice versa, integrating elements of $\N_A$ according to a probability measure on $\s(A)$ yields a state on $\N_A$. 

Therefore, a conditional independence between random variables $A$ and $B$ given the set $\C$
\begin{eqnarray}
p(A \wedge B \, | \, \C) = p(A \, | \, \C) \, p(B \, | \, \C)
\end{eqnarray}
can be rewritten as follows: for any projection $A \in \N_A$, $B \in \N_B$ and $C \in \N_\C$:
\begin{eqnarray}\label{condindc}
\frac{\phi(A \wedge B \wedge C)}{\phi(C)} = \frac{\phi(A \wedge C)}{\phi(C)}\frac{\phi(B \wedge C)}{\phi(C)}
\end{eqnarray}

Although in this paper we stay at the classical level, the theory of von Neumann algebras is wide enough to incorporate also quantum physics. In this case the von Neumann algebras are nonabelian. The events, just like in the classical case, are represented by projections of the von Neumann algebras. In the quantum case conditional independence between the projection $A \in \N_A$ and $B \in \N_B$ given $C \in \N_\C$ reads as follows:
\begin{eqnarray}\label{condindq}
\frac{\phi(CABC)}{\phi(C)} = \frac{\phi(CAC)}{\phi(C)}\frac{\phi(CBC)}{\phi(C)}
\end{eqnarray}
which in the classical case reduces to (\ref{condindc}).

The last point in converting the formalism of Bayesian networks into physics, is to swap the causal graph for spacetime. We can then replace the causal relations embodied in the causal graph by spatiotemporal relations of a given spacetime. Instead of saying that a random variable is the \textit{ancestor} of another variable we will then say that an event is in the \textit{past} of the other. But to do so first we need to \textit{localize} events in spacetime that is we need to have an association of algebras of events to spacetime regions. Such a principled association is offered by the formalism of algebraic quantum field theory. Hence, in the next section we will introduce some elements of algebraic quantum field theory which is indispensable for our purpose which is to come up with a mathematically precise definition of Bell's notion of local causality.

\section{Bell's local causality in a local physical theory}

Let $\ST$ be a globally hyperbolic spacetime and let $\SDK$ be a covering collection of bounded, globally hyperbolic subspacetime regions of $\ST$ such that $(\SDK,\subseteq)$ is a directed poset under inclusion $\subseteq$. A \textit{local physical theory} is a net $\{\OA(V),V\in\SDK\}$ associating algebras of events to spacetime regions which satisfies \textit{isotony} and \textit{microcausality} defined as follows (Haag, 1992; Halvorson 2007; Hofer-Szabó and Vecsernyés 2015, 2016):

\noi
\textit{Isotony}. The net of local observables is given by the isotone map $\SDK\ni V\mapsto\OA(V)$ to unital $C^*$-algebras, that is $V_1 \subseteq V_2$ implies that $\OA(V_1)$ is a unital $C^*$-subalgebra of $\OA(V_2)$. The \textit{quasilocal algebra} $\OA$ is defined to be the inductive limit $C^*$-algebra of the net $\{\OA(V),V\in\SDK\}$ of local $C^*$-algebras.

\noi
\textit{Microcausality}: $\OA(V')'\cap\OA \supseteq \OA(V),V\in\SDK$, where primes denote spacelike complement and algebra commutant, respectively.

\noi
If the quasilocal algebra $\OA$ of the local physical theory is commutative, we speak about a \textit{local classical theory}; if $\OA$ is noncommutative, we speak about a \textit{local quantum theory}. For local classical theories microcausality fulfills trivially.

Given a state $\phi$ on the quasilocal algebra $\OA$, the corresponding GNS representation $\pi_{\phi}\colon\OA\to \mathcal{B}(\mathcal{H}_\phi)$ converts the net of $C^*$-algebras into a net of $C^*$-subalgebras of $B(\mathcal{H}_\phi)$. Closing these subalgebras in the weak topology one arrives at a net of local von Neumann observable algebras: $\N(V):=\pi_{\phi}(\OA(V))'', V\in\SDK$. The \textit{net $\{\N(V),V\in\SDK\}$ of local von Neumann algebras} also obeys isotony and microcausality, hence we can also refer to it as a local physical theory. 

Given a local physical theory, we can turn now to the definition of Bell's notion of local causality. Recall that according to Bell a theory is locally causal if any superluminal correlation is screened-off by a ``full specification of local beables in a space-time region $V_C$'' as shown in Fig.\ \ref{dsep1}. As indicated in the Introduction we need to address three questions. What are ``local beables''? What is ``full specification''? Which are the shielder-off regions? The brief answer to the first two questions is the following. In a local physical theory a ``local beable'' in a region $V$ is an \textit{element} of the local von Neumann algebra $\N(V)$. A ``full specification'' of local beables in region $V$ is an \textit{atomic element} of the local von Neumann algebra $\N(V)$. In this paper we do not comment on these two answers. For a more thoroughgoing discussion on why we think this to be the correct translation of Bell's intuition into our framework see (Hofer-Szabó and Vecsernyés, 2015, 2016). 

\begin{figure}[ht]
\centerline{\resizebox{14cm}{!}{\includegraphics{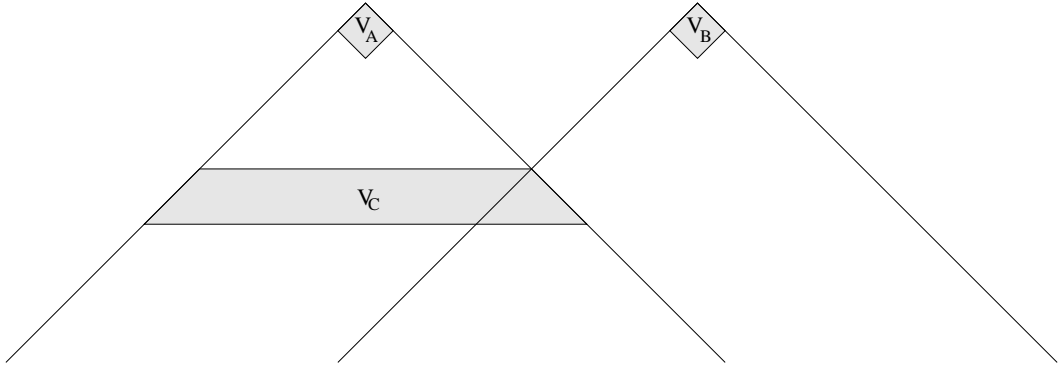}}}
\caption{A completely shielding-off region $V_C$ intersecting with the common past of $V_A$ and $V_B$.}
\label{dsep3}
\end{figure}

To the third question, which is the topic of our paper, the answer is this: a shielder-off region $V_C$ is a region in the causal past of $V_A$ which can block any causal influence on $V_A$ arriving from the common past of $V_A$ and $V_B$. But there is an ambiguity in this answer. Bell's Fig.\ \ref{dsep1} suggests that a shielder-off region should not intersect with the common past. Whereas the requirement of simply blocking causal influences from the past allows for also regions depicted in Fig.\ \ref{dsep3} intersecting with the common past.
This means that one can define a \textit{shielder-off region of $V_A$ relative to $V_B$} either as a region $V_C$ satisfying: 
\begin{enumerate}
\item[]{$\textbf L_1:$} $V_C \subset   J_-(V_A) \qquad (V_C$ is in the causal past of $V_A$),
\item[]{$\textbf L_2:$} $V_A \subset V''_C \qquad (V_C$ is wide enough such that its causal shadow contains $V_A$),
\item[]{$\textbf L_3^Q:$} $V_C\subset V_B' \qquad (V_C$ is spacelike separated from $V_B$)
\end{enumerate}
in tune with Bell's Fig.\ \ref{dsep1}; or one can replace $L_3^Q$ by the weaker requirement
\begin{enumerate}
\item[]{$\textbf L_3^C:$} $J_-(V_C)\supset J_-(V_A) \cap J_-(V_B) \qquad$ (The causal past of $V_C$ contains the common past of $V_A$ and $V_B$)
\end{enumerate}
allowing for regions such as in Fig.\ \ref{dsep2}. It turns out that (with respect to the Bell inequalities, see (Hofer-Szabó and Vecsernyés, 2012b, 2013b)) it is more appropriate to demand $L_3^Q$ in case of a local \textit{quantum} theory and $L_3^C$ in case of a local \textit{classical} theory (hence the superscripts). But note that as the covering regions become infinitely thin shrinking down to a Cauchy surface, requirement $L_3^C$ coincides with requirement $L_3^Q$.

With all these considerations in mind Bell's notion of local causality in the framework of a local physical theory will be the following:

\begin{D}\label{DBLC} A local physical theory represented by a net $\{\N(V),V\in\SDK\}$ of von Neumann algebras is called \textit{locally causal} (in Bell's sense), if \begin{itemize}
\item[(i)] for any pair $A \in\mathcal \N(V_A)$ and $B\in\mathcal \N(V_B)$ of events represented by projections in spacelike separated regions $V_A, V_B\in\SDK$,
\item[(ii)] for every locally normal and faithful state $\phi$ establishing a correlation $\phi(AB)\neq \phi(A)\phi(B)$ between $A$ and $B$,
\item[(iii)] for any spacetime shielder-off region $V_C$ defined by requirements $L_1$, $L_2$ and $L^Q_3/L^C_3$,
\item[(iv)] for any event $C$ in the set $\C$ of atomic events in $\OA(V_C)$,                                                                                                                                                                                                                                                                                                                                                                                                                                                                                                                                                                                                        \end{itemize}
the following screening-off condition holds:
\begin{eqnarray}\label{BLC}
\frac{\phi(CABC)}{\phi(C)} = \frac{\phi(CAC)}{\phi(C)}\frac{\phi(CBC)}{\phi(C)}
\end{eqnarray}
which for a local \textit{classical} theory is equivalent to
\begin{eqnarray}\label{BLC'}
p(A \wedge B \, | \, \C) = p(A \, | \, \C) \, p(B \, | \, \C)
\end{eqnarray} 
\end{D}

\noi
In short, a local physical theory is locally causal in Bell's sense if every superluminal correlation is screened off by all atomic events in all shielder-off region. (For many delicate questions such as what if the algebras are non-atomic, how this definition of local causality relates to the Common Cause Principle and the Bell inequalities see again (Hofer-Szabó and Vecsernyés, 2015, 2016).)

The question left is, however: \textit{why} shielder-off regions are characterized by requirements $L_1$, $L_2$ and $L^Q_3/L^C_3$? To this we turn in the next Section.

\section{Shielder-off regions are d-separating}

The point we are going to make in this Section is that shielder-off regions in the definition of local causality conform to d-separating sets in directed acyclic graphs and to m-separating sets in mixed acyclic graphs. 

First we show how a local physical theory gives rise to a causal graph. Consider a local \textit{classical} theory $\{\N(V),V\in\SDK\}$ where the covering collection is induced by a partition $\T$ of a spacetime $\ST$. By \textit{partition} we mean a countable set of disjoint, bounded spacetime regions such that their union is $\ST$. Whether we demand global hyperbolicity from the elements of the partition will turn out to play an important role in the type of the graph we can construct. For some specific globally hyperbolic coverings we will get directed acyclic graphs, otherwise only a mixed graph.

Let the vertices of the $\G$ be the regions in the partition, $\{V\in \T\}$. Denote the vertex corresponding to the region $V \in \T$ by $A_V$ and the region corresponding to a vertex $A$ by $V_A$. Similarly, denote the \textit{set} of vertices corresponding to the region $V \in \SDK$ by $\C_V$ and the region corresponding to a set of vertices $\C$ by $V_\C$. Define the \textit{ancestors} of a vertex $B$ as:
\begin{eqnarray*}
Anc(B) := \{A \in \V\, | \, A \neq B, \, \, V_A \cap J_-(V_B) \neq \emptyset\}
\end{eqnarray*}
and the \textit{parents} of $B$, $Par(B)$, as those elements in $Anc(B)$ for which there is a causal curve connecting $V_A$ and $V_B$ directly (that is without entering a third region between them). Now, let there be an arrow $A\rightarrow B$ between vertex $A$ and $B$ in $\T$ if and only if $A \in Par(B)$. It will turn out that the type of the graph we obtain is crucially depending on the partition $\T$ of the spacetime. Let us see the different cases.

\begin{figure}[ht]
\centerline{\resizebox{14cm}{!}{\includegraphics{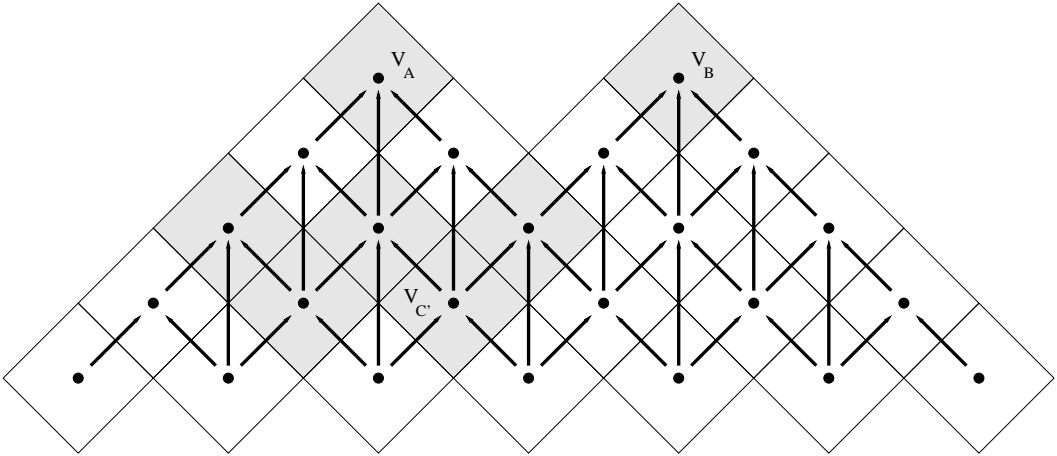}}}
\caption{The directed acyclic graph generated by double cones of equal size covering the 1+1 dimensional Minkowski spacetime.}
\label{dsep6}
\end{figure}

If $\ST$ is the 1+1 dimensional Minkowski spacetime, then it can be covered by double cones of equal size. (See Fig.\ \ref{dsep6}.)
Double cones are globally hyperbolic. (For the details of this example see (Hofer-Szabó, 2015b).) The causal graph corresponding to this covering emerges simply by connecting the midpoints of those adjacent double cones which lie in the causal past of one another. What we get is just the directed acyclic graph depicted in Fig.\ \ref{dsep4} in Section 2.

Fig.\ \ref{dsep6} is a kind of ``superposition'' of a spacetime diagram and a Bayesian network. Consider for example region $V_{\C'}$. Reading Fig.\ \ref{dsep6} as a spacetime diagram, one sees that $V_{\C'}$ is a \textit{shielder-off region} (similar to the one depicted in Fig.\ \ref{dsep3}). Reading Fig.\ \ref{dsep6} as a causal graph, one observes that the set $\C'$ corresponding to $V_{\C'}$ (depicted in Fig.\ \ref{dsep4}) is a \textit{d-separating set}. Similarly, one can check that the region associated to the d-separating set $\C$ in Fig.\ \ref{dsep4} is a shielder-off region and the region associated to the d-connecting set $\C''$ is not a shielder-off region.

\begin{figure}[ht]
\centerline{\resizebox{14cm}{!}{\includegraphics{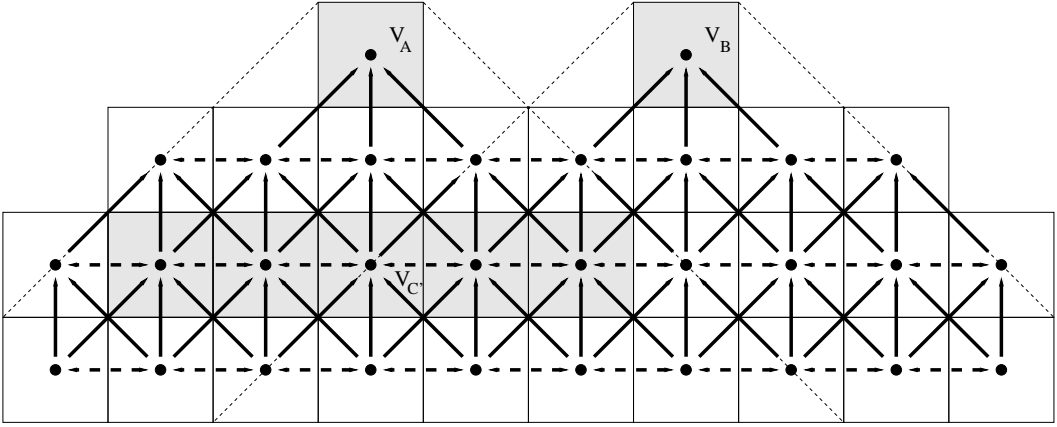}}}
\caption{The mixed acyclic graph generated by boxes of equals size covering of the 1+1 dimensional Minkowski spacetime.}
\label{dsep7}
\end{figure}

A general spacetime $\ST$ cannot be partitioned to globally hyperbolic regions, let alone to double cones. Still one can construct the causal graph corresponding to a partition $\T$. In Fig.\ \ref{dsep7} 
we illustrate such a construction where a 1+1 dimensional Minkowski spacetime is covered by boxes of equals size. (This example, in contrast to the previous one, can be generalized for a $3+1$-dimensional Minkowski spacetime covered by $3+1$-dimensional boxes of equals size.) The causal graph emerging from this construction is not a directed acyclic graph since it contains bi-directed edges: spacelike neighboring boxes will be \textit{spouses}. What we get is a mixed acyclic graph depicted in Fig.\ \ref{dsep5}. Again, confronting Fig.\ \ref{dsep5} and Fig.\ \ref{dsep7} one can see that the set $\C'$ is \textit{not} an m-separating set and at the same time the corresponding region $V_{\C'}$ is \textit{not} a shielder-off region of $V_A$ relative to $V_B$.

\noi
The exact characterization of the graphs emerging from a different coverings of a given spacetime is a subtle question which we do not go into here. Instead we turn now to the construction of random variables. Let $\N(V)$ be the local von Neumann algebra associated to the spacetime region $V \in \T$. Denote by $\s(V)$ the sigma-algebra of the projections of $\N(V)$. Let the random variable (also denoted by) $A_V$ associated to $V$ be any Borel-measurable function from $\s(V)$ to $\B(\RE)$. Any state $\phi$ will then define a probability measure $p$ on $\s(V)$ for any $V \in \T$ and, due to isotony of the net, also for any $V$ which is a  \textit{finite} union of regions in $\T$. (Note that $\s(\ST)$ may not be a sigma-algebra since the quasilocal algebra $\A$ is not necessarily a von Neumann algebra, so it may not contain projections.)

In sum, any finite set of regions of a local classical theory $\{\N(V),V\in\SDK\}$ generated by a globally hyperbolic partition of $\ST$ defines a Bayesian network $(\G, \V)$. If global hyperbolicity is not required, then $\G$ is not a directed acyclic but only a mixed graph.

Now, we state and prove the main claim of the paper.
\begin{Prop}\label{Prop1}
Let $G$ be a directed/mixed acyclic graph constructed from a local classical theory $\{\N(V),V\in\SDK\}$ where $\SDK$ is generated by a partition $\T$ of $\ST$. Suppose that $\{\N(V),V\in\SDK\}$ is locally causal in the sense of Definition \ref{DBLC}. Then for any shielder-off region $V$ defined by $L_1$, $L_2$ and $L^C_3$, the corresponding set $\C_V$ is d-separating/m-separating.
\end{Prop}

\noindent
\textbf{Proof.} Let $A$ and $B$ two collateral vertices in $\G$ corresponding to two spacelike separated regions $V_A$ and $V_B$, respectively ($V_A, V_B \in \T$). Call a set $\C$ of random variables a \textit{shielder-off set} (for $A$ relative to $B$), if $V_\C$ is a shielder-off region (for $V_A$ relative to $V_B$). Shielder-off sets block every directed path from $Anc(A) \wedge Anc(B)$, the set of common ancestors of $A$ and $B$, to $A$ (that is every directed path has to pass through $\C$). 

We show that shielder-off sets are d-separating/m-separating. Let $\C$ be a shielder-off set for $A$ relative to $B$. We have to show that $\C$ blocks every path connecting $A$ and $B$. First consider those paths that contain no colliders. These paths need to pass through the set of common ancestors of $A$ and $B$, $Anc(A) \wedge Anc(B)$. Hence, the shielder-off set $\C$ blocks them. So there remain only those paths to be blocked which contain at least one collider. It is easy to see that these latter paths need to contain at least one collider $E$ such that $E \notin Anc(A)$. But then neither $E$ nor any descendant of $E$ is in $\C$, hence $\C$ blocks also these paths. \qed\

\begin{figure}[ht]
\centerline{\resizebox{14cm}{!}{\includegraphics{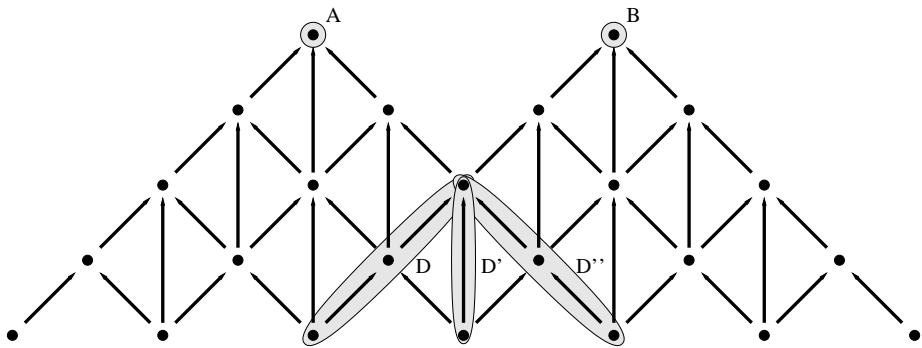}}}
\caption{Minimal d-separating but not shielder-off regions.}
\label{dsep9}
\end{figure}

\noi
The converse of Proposition \ref{Prop1} is not true: d-separating sets are not necessarily shielder-off sets. Tian, Paz, and Pearl (1998) list algorithms to find the so-called \textit{minimal d-separating sets} for two random variables $A$ and $B$, that is sets that are d-separating but taking away any vertex from the set they will cease to be d-separating. It turns out that any minimal d-separating set is sitting in the \textit{union} of the ancestors of $A$ and $B$ (including also $A$ and $B$), $Anc(A) \vee Anc(B) \vee A \vee B$. However, a minimal d-separating set need not satisfy relations $L_1$, $L_2$ and $L^C_3$. For example the sets $\mathcal D$, $\mathcal D'$ and $\mathcal D''$ in Fig.\ \ref{dsep9} are all minimal d-separating sets but not shielder-off regions for $A$ relative to $B$. 

At any event, shielder-off regions are d-separating, and this was to be shown in this paper.

\section{Conclusions}

The aim of the paper was to motivate Bell's definition of local causality by means of Bayesian networks. To this aim, first we constructed a causal graph from the covering collection of a spacetime. In certain cases the graph was a directed acyclic graph, in other cases only a mixed acyclic graph. Similarly, we have associated random variables to the local algebras of a local physical theory. By this move shielder-off regions turned out be specific d-separation (m-separating) sets on the causal graph. Hence, Bell's definition of local causality requiring that spacelike separated events should be screened-off by events in a shielder-off region turned out to be a d-separation criterion. 
\vspace{0.2in}

\noindent
{\bf Acknowledgements.} I wish to thank Péter Vecsernyés for valuable discussions. This work has been supported by the Hungarian Scientific Research Fund, OTKA K-100715 and OTKA K-115593 and by the Bilateral Mobility Grant of the Hungarian and Polish Academies of Sciences, NM-104/2014.

\section*{References} 
\begin{small}
\begin{list} 
{ }{\setlength{\itemindent}{-15pt}
\setlength{\leftmargin}{15pt}}

\item J.S. Bell, ''La nouvelle cuisine,'' in: J. Sarlemijn and P. Kroes (eds.), {\it Between Science and Technology}, Elsevier, (1990); reprinted in (Bell, 2004, 232-248).

\item J.S. Bell, \textit{Speakable and Unspeakable in Quantum Mechanics}, (Cambridge: Cambridge University Press, 2004).

\item J. Butterfield, ''Stochastic Einstein Locality Revisited,''  \textit{Brit. J. Phil. Sci.}, \textbf{58},  805-867, (2007).

\item J. Earman and G. Valente,  ''Relativistic causality in algebraic quantum field theory,'' \textit{Int. Stud. Phil. Sci.}, \textbf{28 (1)}, 1-48 (2014).

\item C. Glymour, ''Markov properties and quantum experiments,'' in W. Demopoulos and I. Pitowsky (eds.) \textit{Physical Theory and its Interpretation}, (Springer, 117-126, 2006).

\item C. Glymour, R. Scheines and P. Spirtes, ''Causation, Prediction, and Search,'' (Cambridge: The MIT Press , 2000).

\item R. Haag, \textit{Local quantum physics}, (Heidelberg: Springer Verlag, 1992).

\item H. Halvorson, ''Algebraic quantum field theory,'' in J. Butterfield, J. Earman (eds.), \textit{Philosophy of Physics, Vol. I}, Elsevier, Amsterdam, 731-922 (2007).

\item J. Henson, ''Comparing causality principles,'' \textit{Stud. Hist. Phil. Mod. Phys.}, \textbf{36}, 519-543 (2005).

\item J. Henson, ''Confounding causality principles: Comment on Rédei and San Pedro's ``Distinguishing causality principles'','' \textit{Stud. Hist. Phil. Mod. Phys.}, \textbf{44}, 17-19 (2013a).

\item J. Henson, ''Non-separability does not relieve the problem of Bell's theorem,'' \textit{Found. Phys.}, \textbf{43}, 1008-1038 (2013b).

\item G. Hofer-Szabó, M. Rédei and L. E. Szabó, \textit{The Principle of the Common Cause}, (Cambridge: Cambridge University Press, 2013).

\item G. Hofer-Szabó and P. Vecsernyés, ''Reichenbach's Common Cause Principle in AQFT with locally finite degrees of freedom,'' \textit{Found. Phys.}, \textbf{42}, 241-255 (2012a).

\item G. Hofer-Szabó and P. Vecsernyés, ''Noncommuting local common causes for correlations violating the Clauser--Horne inequality,'' \textit{J. Math. Phys.}, \textbf{53}, 12230 (2012b).

\item G. Hofer-Szabó and P. Vecsernyés, ''Noncommutative Common Cause Principles in AQFT,'' \textit{J. Math. Phys.}, \textbf{54}, 042301 (2013a).

\item G. Hofer-Szabó and P. Vecsernyés, ''Bell inequality and common causal explanation in algebraic quantum field theory,'' \textit{Stud. Hist. Phil. Mod. Phys.},  \textbf{44 (4)}, 404–416 (2013b).

\item G. Hofer-Szabó and P. Vecsernyés, ''On the concept of Bell's local causality in local classical and quantum theory,''  \textit{J. Math. Phys}, \textbf{56}, 032303 (2015)

\item G. Hofer-Szabó and P. Vecsernyés, ''A generalized definition of Bell's local causality,'' \textit{Synthese} (forthcoming) (2016).

\item G. Hofer-Szabó, ''Local causality and complete specification: a reply to Seevinck and Uffink,'' in U. M{\"a}ki, I. Votsis, S. Ruphy and G. Schurz (eds.) \textit{Recent Developments in the Philosophy of Science: EPSA13 Helsinki}, Springer Verlag, 209-226 (2015a).

\item G. Hofer-Szabó, ''Relating Bell's local causality to the Causal Markov Condition,''  \textit{Found. Phys.}, \textbf{45(9)}, 1110-1136 (2015b).

\item T. Maudlin, ``What Bell did,'' \textit{J. Phys. A: Math. Theor.}, \textbf{47}, 424010 (2014).

\item T. Norsen, ''Local causality and Completeness: Bell vs. Jarrett,'' \textit{Found. Phys.}, \textbf{39}, 273 (2009).

\item T. Norsen, ''J.S. Bell's concept of local causality,'' \textit{Am. J. Phys}, \textbf{79}, 12, (2011).

\item J. Tian, A. Paz, and J. Pearl, ``Finding minimal d-separating sets,'' \textit{UCLA Cognitive Systems Laboratory, Technical Report} (R-254), (1998).

\item J. Pearl, ''Causality: Models, Reasoning, and Inference,'' Cambridge: (Cambridge University Press, 2000).

\item M. Rédei, ''Reichenbach's Common Cause Principle and quantum field theory,'' \textit{Found. Phys.}, \textbf{27}, 1309-1321 (1997).

\item M. Rédei, ''A categorial approach to relativistic locality,'' \textit{Stud. Hist. Phil. Mod. Phys.}, \textbf{48}, 137-146 (2014). 

\item M. Rédei and I. San Pedro, ''Distinguishing causality principles,'' \textit{Stud. Hist. Phil. Mod. Phys.}, \textbf{43}, 84-89 (2012). 

\item M. Rédei and J. S. Summers, ''Local primitive causality and the Common Cause Principle in quantum field theory,'' \textit{Found. Phys.}, \textbf{32}, 335-355 (2002).

\item T. S. Richardson and P. Spirtes, ''Ancestral graph Markov models,'' \textit{Ann. Statist.} \textbf{30}, 962–1030 (2002).

\item K. Sadeghi and S. Lauritzen, ''Markov properties for mixed graphs,'' \textit{Bernoulli.} \textbf{20/2}, 676-696 (2014).

\item M. P. Seevinck and J. Uffink, ''Not throwing our the baby with the bathwater: Bell's condition of local causality mathematically 'sharp and clean', '' in: Dieks, D.; Gonzalez, W.J.; Hartmann, S.; Uebel, Th.; Weber, M. (eds.) \textit{Explanation, Prediction, and Confirmation The Philosophy of Science in a European Perspective}, Volume 2, 425-450 (2011).

\item M. Suárez and I. San Pedro ''Causal Markov, robustness and the quantum correlations,'' in M. Suarez (ed.), \textit{Probabilities, causes and propensities in physics}, 173-193. Synthese Library, 347, (Dordrecht: Springer, 2011)

\item M. Suárez, ''Interventions and causality in quantum mechanics,'' \textit{Erkenntnis}, \textbf{78}, 199-213 (2013).

\end{list}
\end{small}

\end{document}